\documentclass{moriond}

\def\Journal#1#2#3#4{{#1} {\bf #2}, #3 (#4)}

\def\be{\begin{equation}}
\def\ee{\end{equation}}
\def\bea{\begin{eqnarray}}
\def\eea{\end{eqnarray}}

\usepackage{amsmath} 
\usepackage{amssymb}
\newcommand{\diff}{{\rm d}}

\newcommand{\Mhs}{\ensuremath{M_{200}^{\rm{HS}}}}

\newcommand{\fbr}{\ensuremath{\mathcal{F}_{\beta}(r)}}
\newcommand{\M}{\ensuremath{M_{500}}}
\newcommand{\btwo}{\ensuremath{\beta_{2}}}
\newcommand{\phitwo}{\ensuremath{\phi_{\infty, 2}}}
\newcommand{\fofr}{\text{$f(R)$} }

\newcommand{\GNeff}{\ensuremath{G_{\rm N}^{\rm eff}}}
\newcommand{\gNtilde}{\ensuremath{\tilde{\gamma}_{\rm N}}}

\newcommand{\GN}{\ensuremath{G_{\rm N}}}
\newcommand{\Mcau}{\ensuremath{M_{200}^{\rm{Cau}}}}

\newcommand{\Photo}{\includegraphics[height=35mm]{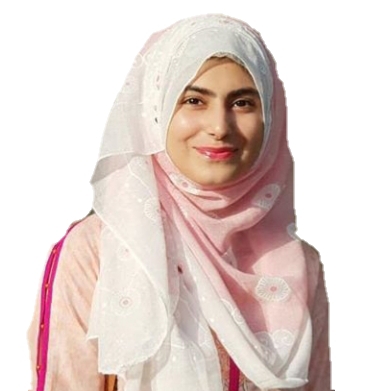}}

\begin{document}
\title{ Modified gravity in galaxy clusters: Joint analysis of Hydrostatics and Caustics }

\author{Minahil Adil Butt$^{1,2}$, Sandeep Haridasu$^{1,2,3}$, Yacer Boumechta$^{1,2}$, Francesco Benetti$^{1,2}$, \\ Lorenzo Pizzuti$^{4}$, Carlo Baccigalupi$^{1,2,3}$, Andrea Lapi$^{1,2,3,5}$}

\address{$^1$SISSA, Via Bonomea, 265, Trieste, Italy \\ 
$^2$INFN, Sezione di Trieste, Via Valerio 2, I-34127 Trieste, Italy\\
$^3$IFPU, Institute for Fundamental Physics of the Universe, via Beirut 2, 34151 Trieste, Italy \\
$^4$Universit\'a degli Studi di Milano Bicocca, Piazza della Scienza 3, I-20126 Milano, Italy \\
$^5$IRA-INAF, Via Gobetti 101, 40129 Bologna, Italy
}

\maketitle

\abstracts{
We present a comprehensive joint analysis of two distinct methodologies for measuring the mass of galaxy clusters: hydrostatic measurements and caustic techniques. We show that by including cluster-specific assumptions obtained from hydrostatic measurements in the caustic method, the potential mass bias between these approaches can be significantly reduced. Applying this approach to two well-observed massive galaxy clusters A2029 and A2142. We find no discernible mass bias, affirming the method's validity. We then extend the analysis to modified gravity models and draw a similar conclusion when applying our approach. Specifically, our implementation allows us to investigate Chameleon and Vainshtein screening mechanisms, tightening the posteriors and enhancing our understanding of these modified gravity scenarios.}

\section{Introduction}
\label{subsec:prod}
The caustic technique \cite{Diaferio99,Diaferio_1997}, provides a unique approach to estimating the mass of the clusters, utilizing the escape velocity profile of member galaxies within the cluster. The caustic surface defines the boundaries of the projected phase space. The amplitude of these caustics, denoted as $\mathcal{A}(r)$, decreases as we move away from the cluster center, and it is tied to the average velocity component $\langle v^{2}\rangle$\cite{Diaferio_2005}. In a spherically symmetric system, the escape velocity $v_{\rm{esc}}^{2}(r)$ is directly related to the gravitational potential $\Phi(r)$ and is a non-increasing function of radial distance.\\
To quantify this average velocity component, the velocity anisotropy profile $\beta(r)$ is utilized. This profile $\beta(r)$ is expressed as $1-(\langle v_{\theta}^{2}\rangle+ \langle v_{\phi}^{2}\rangle)/2\langle v_{r}^{2}\rangle$, where $v_{\theta}$, $v_{\phi}$, and $v_{r}$ are the longitudinal, azimuthal, and radial velocity components of galaxies, respectively. The gravitational potential profile is related to the caustic amplitude through the function $g(\beta)$, given by:
\begin{equation}
    g(\beta)=\frac{3-2\beta(r)}{1-\beta(r)}.
    \label{eqn:gbeta}
\end{equation}
After estimating the caustic amplitude, $\beta(r)$ becomes the sole unknown factor in estimating the gravitational potential. 
\begin{equation}
    \label{eqn:phi_caustic}
        -2\Phi(r)=\mathcal{A}^{2}(r)g(\beta).
    \end{equation}
The equation for the caustic mass profile of a spherical system can be expressed as:
\begin{equation}
        \label{eqn:Mass_Profile}
        GM(<r)=\int_{0}^{r}\mathcal{A}^{2}(r)\mathcal{F}_{\beta}(r)dr
\end{equation}
where $\mathcal{F}_{\beta}(r) =\mathcal{F}(r)g(\beta)$ and 
\begin{equation}
    \label{eqn:fofr}
    \mathcal{F}(r)=-2\pi G\frac{\rho(r)r^{2}}{\Phi(r)}.
\end{equation}
In this context, $\rho(r)$ is the density profile of the spherical system (which we assume to be the NFW profile, leaving the investigation with other mass models to a future analysis) and $\Phi(r)$ stands for the gravitational potential profile. \\

\subsection{Modified Gravity Models}
\label{sec:MG}

\textit{Chameleon Screening}: The chameleon model \cite{Khoury:2003aq} modifies gravity by introducing a scalar field non-minimally coupled with the matter components and gives rise to a fifth force that can be of the same order as the standard gravitational force. The chameleon mechanism is a two-parameter model. The chameleon model parameter $\beta$ determines the strength of the fifth force when it is not screened. The second chameleon parameter, $\phi_{\infty}$, controls the effectiveness of the screening mechanism. The chameleon mechanism operates whenever a scalar field couples to matter in such a way that its effective mass depends on the local matter density. The theory (in the weak-field limit and for the non-relativistic matter) is given by \cite{Terukina:2013eqa}:
\begin{equation}
    L_{\mathrm{chameleon}}=-\frac{1}{2}(\partial\phi)^{2}-V(\phi)-\frac{g_{\rm c}\phi}{M_{\mathrm{Pl}}}\rho_{\mathrm{m}}
\end{equation}
The dimensionless coupling parameter $g_{\rm c}$ is assumed to be $O(1)$, corresponding to the gravitational strength coupling.\\
The modified gravitational potential under this mechanism is given by \cite{Wilcox:2015kna}:

\begin{equation}
    \label{eqn:grav_cham}
    \frac{ \diff { \Phi(r)}}{\diff {r}}=\frac{G_{\mathrm{N}}{M}(r)}{r^{2}} + \beta \frac{\diff{\phi}}{\diff {r}},
\end{equation}
where $\phi(r)$ is the chameleon field.

\textit{Vainshtein Screening}: In the case of Vainshtein screening the additional degrees of freedom are screened through a non-linear mechanism. In Vainshtein screening the gravitational potentials are modified inside the matter sources. These modifications are screened outside the sources and GR is recovered in low-density environments, as required by tests of gravity. GR is not recovered everywhere within the Vainshtein radius as the screening breaks down there. The Vainshtein radius $r_{V}$ is given by the curvature of the object \cite{Paillas_2019}. The modified gravitational potential is now given as \cite{Cardone2020Aug,Langlois_2017},
\begin{equation}
\label{eqn:grav_vain}
    \frac{\diff \Phi(r)}{\diff r}=\frac{G_{\mathrm{N}}^{\mathrm{eff}}{M}(r)}{r^{2}}+\Xi_{1}G_{\mathrm{N}}^{\mathrm{eff}}{M}''(r),
\end{equation}
where $G_{\mathrm{N}}^{\mathrm{eff}}$ is the modified Newton's constant defined as $G_{\mathrm{N}}^{\mathrm{eff}}=\tilde{\gamma}_{\mathrm{N}} \times \GN$ and $'$ represents the derivative w.r.t $r$. 

\section{Conclusion}

\begin{figure*}
    \centering
    \includegraphics[scale=0.285]{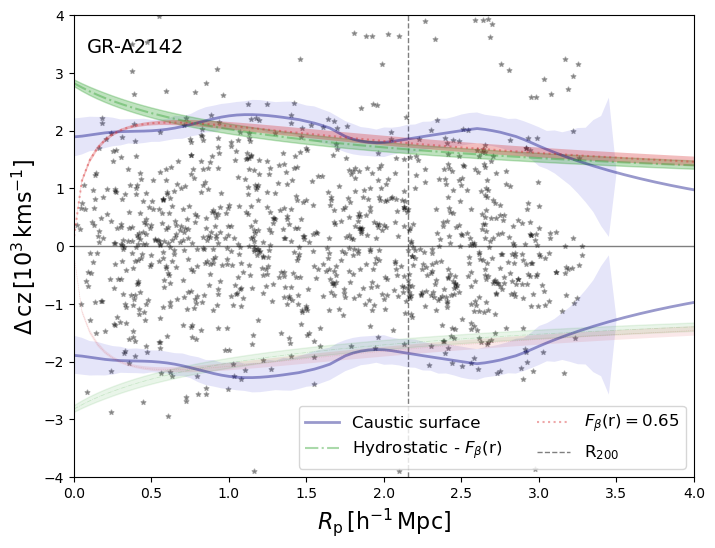}
    \includegraphics[scale=0.285]{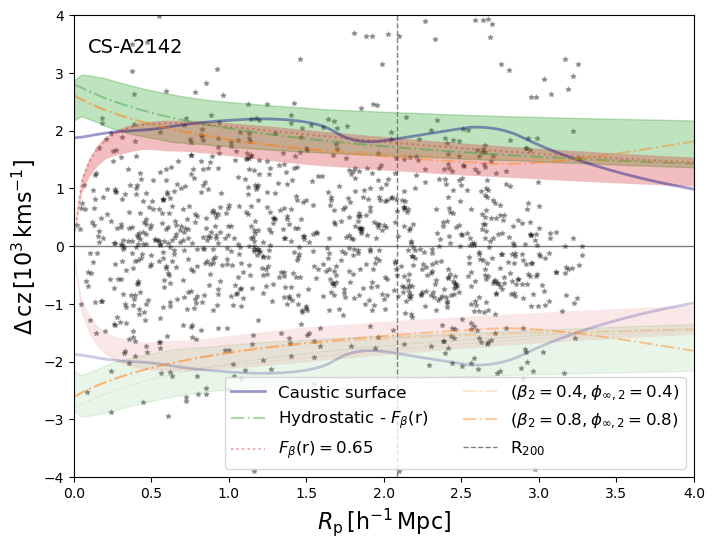}
    \includegraphics[scale=0.285]{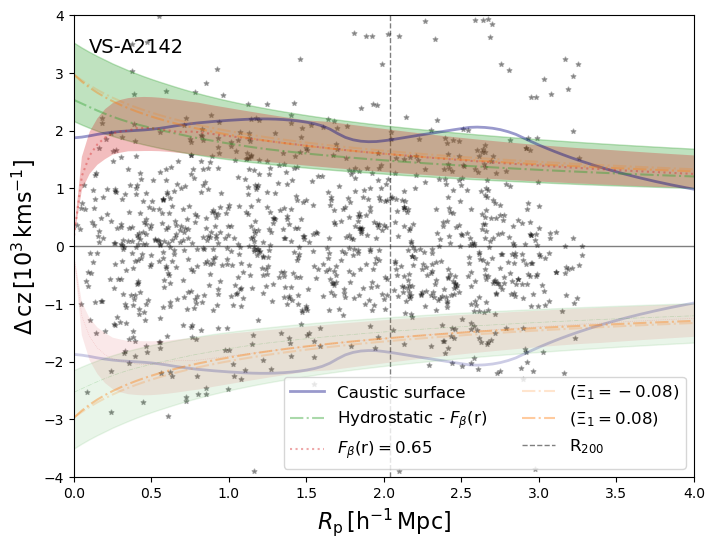}
    \caption{Caustic Profiles for the cluster A2142. We show the caustic profile and the corresponding uncertainty in shaded blue (in \textit{left}) regions, alongside the phase space distribution of the galaxies. \textit{Left}: We compare the caustic profiles estimated using the hydrostatic data in the case of GR, shown in green. {The caustic profile estimated using the hydrostatic data and a constant $\fbr = 0.5$ are shown as well.} \textit{Center}: Same as the \textit{Left} panel comparing against the caustic surface estimated using the hydrostatic data in the case of chameleon screening. \textit{Right}: Same as the other two panels, comparing the caustic surface in the case of Vainshtein screening. {For the Chameleon and Vainshtein screening cases, we show a few expected caustic surfaces for various values of assumed values of the modified gravity parameters.}}
    \label{fig:causticprofiles-A2142}
\end{figure*}

We begin by presenting our results for the construction of the caustic profiles for cluster A2142 considered in our analysis \footnote{For brevity, we have omitted the analysis of A2029 in this proceeding. Please see the paper\cite{Butt:2024jes} for details.}. Figure \ref{fig:causticprofiles-A2142} displays the reconstructed caustic surface (blue) alongside the caustic surface predicted by the hydrostatic technique\cite{Ettori_2019} assuming an NFW mass profile (green). We observe good agreement between the two estimates. In the \textit{Center} and the \textit{Right} panels we show the same comparison against the Chameleon and Vainshtein screening, respectively. In particular, we note a sharp deviation in the caustic surface predicted by hydrostatic equilibrium at the screening radius in the case of Chameleon screening, indicating constraints on certain parameter space. For Vainshtein screening, modifications extend beyond the screening radius, affecting the inner regions of the galaxy clusters. We begin by contrasting our caustic mass estimates to those present in literature to validate our procedure assuming a constant $\fbr$.

\begin{figure*}
    \centering
    \includegraphics[scale=0.6]{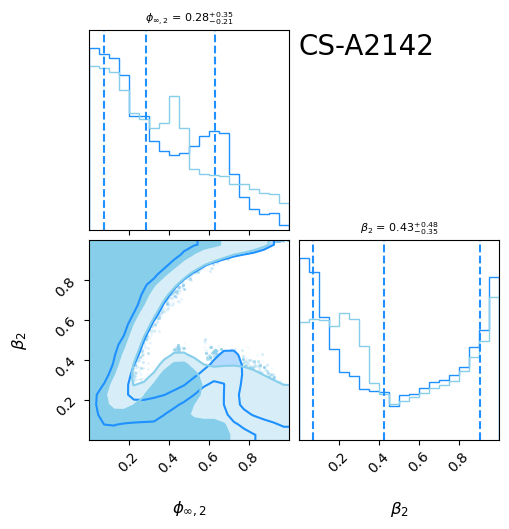}
    \includegraphics[scale=0.6]{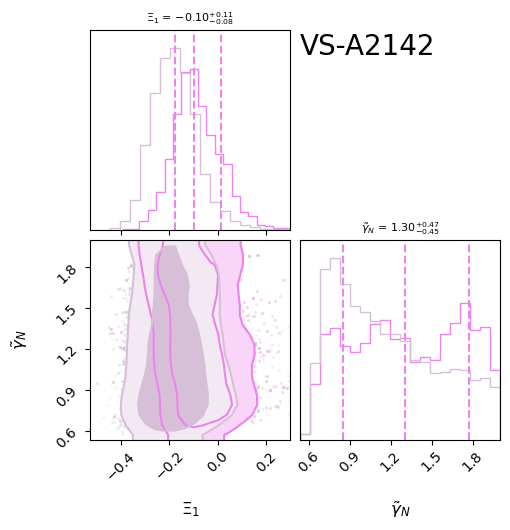}
    \caption{\textit{Left}: We show the $68\%$ and $95\%$ C.L. contours in the case of the \textit{Chameleon screening} (CS) using the galaxy cluster A2142. The {\textit{light blue}} filled contours show the constraints obtained using only the hydrostatic data. {Similarly, the contours outlined in dark blue} show the joint analysis including the caustic technique. \textit{Right}: Same as \textit{Left} but for \textit{Vainshtein screening} (VS).} 
    \label{fig:MGContours-A2142}
\end{figure*}

\textit{Chameleon screening}: 
As can be seen in Figure \ref{fig:MGContours-A2142}, the joint analysis of the hydrostatic and caustic techniques helps reduce the degenerate region between mass ($\M$) and the coupling constant ($\btwo$) parameter, which we earlier elaborated upon\cite{Boumechta:2023qhd}. Traditionally, the approach to alleviate this degeneracy and obtain constraints was to assume a mass prior obtained using weak lensing techniques and we introduced an internal mass prior\cite{Boumechta:2023qhd} which is a prior of the mass-concentration obtained when restricting the analysis to $\btwo\geq 0.5$. In the current work, we utilize the Caustic technique to fulfill the role of a prior, however, the formalism is not to utilize any mass prior but to perform an importance sampling while simultaneously constraining the mass from the caustic technique. The ability of the Caustic technique to improve the constraints heavily relies on the mass estimates alone. 
Additionally, Figure \ref{fig:MGContours-A2142} shows that the constraint on $\phitwo$ for the case of $\btwo = \sqrt{1/6}$, which corresponds to the specific case of $\fofr$ gravity is largely reduced. 
This improvement here is essentially driven by the caustic mass we constrain taking into account the $\fbr$ already constrained using the hydrostatic data is much tighter than the WL masses\cite{Herbonnet:2019byy} which we have earlier utilized\cite{Boumechta:2023qhd}.

\textit{Vainshtein screening}: As described in \ref{sec:MG}, within the Vainshtein screening the gravitational potential and the weak lensing potential are distinct and a joint analysis would be needed to correctly constrain the running of the gravitational constant $\GNeff$. As the $\fbr$ necessary to constrain the caustic mass is the gravitational potential as well, we do not immediately break the degeneracy between the $\gNtilde$ and $\Mhs$. The current cluster A2142 is a very good example of a cluster deviating mildly from GR (We observe a situation closer to GR with A2029). 
Cluster A2142, which shows a mild $\sim 2\sigma$ deviation with $\Xi_1 \sim -0.20^{+0.10}_{-0.08}$ using only the hydrostatic data, is now more consistent with GR within $\sim 1\sigma$, having $\Xi_1 = -0.10^{+0.11}_{-0.06} $. Note that this is a shift in the parameters with no major improvement in the relative constraint. This improvement in the agreement with the GR is essentially because the masses estimated using the hydrostatic method and the caustic method have a bias of $\Mhs/ \Mcau = 0.81^{+0.09}_{-0.07}$, which is about the same significance $\geq 2\sigma$, as the earlier for modified gravity using the hydrostatic data alone. Nevertheless, we perform the joint analysis as the inconsistency is only of the order of $\sim 2 \sigma$. 

\newpage
\section*{Acknowledgments}

AL has been supported by the EU H2020-MSCA-ITN-2019 Project 860744 `BiD4BESt: Big Data applications for Black Hole Evolution Studies' and by the PRIN MIUR 2017 prot. 20173ML3WW, `Opening the ALMA window on the cosmic evolution of gas, stars, and supermassive black holes'. BSH is supported by the INFN INDARK grant and acknowledges support from the COSMOS project of the Italian Space Agency (cosmosnet.it). CB acknowledges support from the COSMOS project of the Italian Space Agency (cosmosnet.it), and the INDARK Initiative of the INFN (web.infn.it/CSN4/IS/Linea5/InDark).

\end{document}